\documentclass[a4paper,10pt,twoside]{cpc-hepnp}

\usepackage{multicol}
\usepackage{graphicx}
\usepackage{booktabs}
\usepackage{amssymb,bm,mathrsfs,bbm,amscd}
\usepackage[tbtags]{amsmath}
\usepackage{lastpage}
\usepackage{CJK}

\begin{document}
\begin{CJK*}{GBK}{song}

\fancyhead[c]{\small Chinese Physics C~~~Vol. XX, No. X (201X)
XXXXXX} \fancyfoot[C]{\small 010201-\thepage}

\footnotetext[0]{Received 14 March 2009}

\title{Spontaneous Magnetization of Solid Quark-cluster Stars\thanks{Supported by the 973 program (No. 2012CB821801),
the National Natural Science Foundation of China (11203018, 11225314, 11365022),
the West Light Foundation (XBBS-2014-23),
the Science Project of Universities in Xinjiang (XJEDU2012S02) and
the Doctoral Science Foundation of Xinjiang University (BS120107). }}

\author{%
      LAI Xiao-Yu$^{1,2;1)}$\email{laixy@pku.edu.cn}%
\quad XU Ren-Xin$^{3,4)}$
}
\maketitle

\address{%
$^1$ Xinjiang Astronomical Observatory, Chinese Academy of Sciences, Urumqi 830011, China\\
$^2$ School of Physics, Xinjiang University, Urumqi 830046, China\\
$^3$ School of Physics, Peking University, Beijing 100871, China\\
$^4$ Kavli Institute for Astronomy and Astrophysics, Peking University, Beijing 100871, China\\
}

\begin{abstract}
Pulsar-like compact stars usually have strong magnetic fields, with the strength
from $\sim 10^8$ to $\sim 10^{12}$ Gauss on surface.
How such strong magnetic fields can be generated and maintained is still an unsolved problem, which is, in
principle, related to the interior structure of compact stars, i.e., the equation of state of cold matter at supra-nuclear density.
In this paper we are trying to solve the problem in the regime of solid quark-cluster stars.
Inside quark-cluster stars, the extremely low ratio of number density of electrons to that of baryons $n_e/n_b$ and the screening effect from quark-clusters could reduce the long-range Coulomb interaction between electrons to short-range interaction.
In this case, the Stoner's model could apply, and we find that the condition for ferromagnetism is consistent with that for validity of Stoner's model.
Under the screened Coulomb repulsion, the electrons inside the stars could spontaneously magnetized and become ferromagnetic, and hence would contribute non-zero
net magnetic momentum to the whole star.
We conclude that, for most cases in solid quark-cluster stars, the amount of net magnetic momentum,
which is proportional to the amount of unbalanced spins $\xi=(n_+-n_-)/n_e$ and depends on the number density of electrons
$n_e=n_++n_-$, could be significant with non-zero $\xi$.
The net magnetic moments of electron system in solid quark-cluster stars
could be large enough to induce the observed  magnetic fields for pulsars with $B\sim 10^{11}$ to $\sim 10^{13}$ Gauss.
%
%This mechanism of generating magnetic field is not valid for so-called magnetars since extra gravitational and elastic energy, rather than magnetic energy, could be released to power anomalous X-ray pulsars and soft gamma-ray repeaters in the solid quark-cluster star model.
\end{abstract}

\begin{keyword}
Pulsars, strange stars, magnetic fields
%keyword,  3--8 words separated by comma
\end{keyword}

\begin{pacs}
26.60.Kp, 71.10.Ca, 75.50.Gg
%1--3 PACS(Physics and Astronomy Classification Scheme, http://www.aip.org/pacs/pacs.html/)
\end{pacs}

\footnotetext[0]{\hspace*{-3mm}\raisebox{0.3ex}{$\scriptstyle\copyright$}2013
Chinese Physical Society and the Institute of High Energy Physics
of the Chinese Academy of Sciences and the Institute
of Modern Physics of the Chinese Academy of Sciences and IOP Publishing Ltd}%

\begin{multicols}{2}

\section{introduction}

The states of matter of pulsar-like compact stars is a long-standing problem,
although the discovery of pulsars
dates back to nearly half a century ago.
Some efforts have been made to understand the nature of pulsars, among which the model of
quark-cluster stars has been proposed.
With a stiff equation of state, the model of quark-cluster stars suggests the exist of high mass
($>2M_\odot$) pulsars~\cite{Lai:2008cw,LX2009b}, to be favored by the discoveries of massive pulsars
~\cite{Demorest:2010bx, Antoniadis2013}.
For traditional models of neutron stars, however, there are two challenges faced: the so called
``hyperon puzzle'' and the quark-deconfinement~\cite{report2014}.
It is worth mentioning that no such kind of embarrassment exists in this quark-cluster star model since
quark-clusters would be hadron-like.
Composed of clustered quarks and solidified at low enough temperatures, the solid quark-cluster stars
could have stiff equation of states naturally, and consequently they could have high masses, as was also demonstrated in the corresponding-state approach to the equation of state~\cite{Guo2014}.
Besides different manifestations~\cite{Xu:2010}, in addition, glitch phenomenon, including both negligible and significant energy releases during glitches,
could also be well understood in the solid quark-cluster star model~\cite{zhou2014}.
In spite of those successes, what about the strong magnetic field in the solid quark-cluster star model?

In fact, the origin of strong magnetic fields of pulsar-like compact stars is also a
long-standing problem.
In the framework of neutron star models,
the simplest and most popular hypothesis is that, the conservation of magnetic flux resulting from the
frozen of magnetic filed to the star's surface magnify the strength of magnetic field  by some
orders of magnitude.
However, the fossil fields could not be adequate because only a very small fraction of the
progenitors have magnetic fields high enough to produce significant fossil fields for pulsars,
and the conservation of magnetic flux seems to be contrary with the high rotation rates of
pulsars~\cite{spruit2008}.
%
%Therefore, the fossil hypothesis is not a convincing solution generally accepted.
%
The large scale magnetic fields generated by dynamo processes are associated with convection,
but the required equipartition field strength is much larger than what a progenitor star can
offer~\cite{spruit2008}.
Moreover, the inherited magnetic fields are also dissipative.
The time scale of Ohmic diffuse of magnetic field for a typical pulsar could be estimated
as~\cite{goldreich1992} $\tau_{\rm ohmic}\sim 2\times 10^{11}\ {\rm yr}.
$
Although the above time scale is large, the Ohmic decay of magnetic fields could lead to
detectable effects.
However, no convincing observational evidence for decay of magnetic fields has been found.

In the solid quark-cluster star model, the origin of large scale magnetic fields should be very different from that of a
fluid neutron star.
The melting temperature of a solid quark-cluster stars could be as high as 10 MeV (approximated by the potential
depth between quark-clusters~\cite{LX2009b}), which is much higher than that of neutron stars.
In the core of the progenitor star, the dynamo processes could play an important role, but after the solidification
the convection would stop and the produced magnetic fields could be weak.
For a rigid body, the dissipation of macroscopic magnetic fields should also be a problem.
On the other hand, if the magnetic field is intrinsically originated via symmetry broken spontaneously,
then there would be no dissipation process.
The magnetic moment of electrons is much larger than that of quark-clusters, so electrons could
significantly contribute magnetic moment to the whole star.
If we take the intrinsic magnetic moment of electrons as the possible elements giving rise to
the macroscopic magnetism,
then it seems to be similar to the case of ferromagnetism of normal material in terrestrial environment.

The ferromagnetism of normal material is studied extensively in condensed matter
physics.
The origin of ferromagnetism is the correlation between electrons under Coulomb interaction.
For quark-cluster stars, the situation is in fact simpler because electrons are not confined to ``nuclei''
(i.e. the quark-clusters in lattices), and all of the electrons are itinerant.
The ferromagnetism of electron gas has been studied~\cite{Walecka}, and it was found that in the high density limit the kinetic energy dominates and the ground state is unpolarized.
This text-book presentation of high density behavior of an electron gas could be quite different from the case we are focus on.

In solid quark-cluster stars,
{\it the screening effect coming from the positively charged and polarized quark-clusters could be significant}, due to the low ratio of number densities of electrons to that of baryons $Y_e=n_e/n_b$, see Section~{\ref{subsection_spin}}.
This leads to the validity of Stoner's model~\cite{Stoner's}, which fails in quantitatively reproduce the experimental observations for normal solid state systems (a full review about Stoner's model and the related progress both in theory and experiments can be found in~\cite{bruun2014} and the references therein).
For the sake of simplicity, we are assuming a $\delta-$ interaction in Eq.(\ref{Vij}) to simulate the screen effect of huge polarized quark-clusters.

A quark-cluster star could serve as an ideal system where the itinerant ferromagnetism might occur under the Stoner's theory.
The Stoner's model considers screened short-range Coulomb interaction, where
the physical picture of ferromagnetism in repulsive Fermi gases can be
understood as the result of the competition between the repulsive interaction
and the Pauli exclusion principle.
For a perfect electron gas, electrons tend to have balanced spins to save kinetic energy.
However, taking into account the interaction between electrons, they tend to have unbalanced spins to save interaction energy.
It should be noticed that the spin-alignment of electrons under Coulomb repulsion is a result which can be found in textbooks.
For example, ref~\cite{huang} (chapter 11.7) gives a physical picture about this mechanism for spontaneous magnetization.
Coulomb repulsion makes a pair of electrons to favor an antisymmetric spatial wave function to lower the interaction energy.
Because the wave function of a pair of electrons is $\psi(1,2)=\psi^{\rm spin}(1,2)\psi^{\rm space}(1,2)$,
an antisymmetric spatial wave function requires a symmetric spin wave function, which means that they form a spin-triplet state.
In this paper, we demonstrate this mechanism quantitatively in the physical conditions of pulsar-like compact stars, and find that
under certain conditions, the interaction energy becomes significant and the ground state of electrons becomes polarized.
Although the amount of unbalanced spins is small compared to the total spins, the net magnetic moment of the electron system
could be high enough to induce the magnetic fields of typical pulsars.

Certainly the state of compact stars is much different from normal matter, although we have not enough
knowledge about the former.
In solid quark-cluster stars, the quarks are localized in clusters by strong interaction,
so they would not make contribution to the total magnetic moment although
they are Fermions.
Although both of the origin of magnetic fields of compact stars and the nature
of compact stars are uncertain to us, we propose here a possible way to solve the former,
and hope it could give us some hints to the latter.
If the electrons in solid quark-cluster star could spontaneously
magnetization and give rise to enough strong magnetic fields,
then we can give some constraints about $Y_e$,
which could reflect some properties of strong interactions.

Defining the amount of unbalanced spins
\begin{equation}
\xi=\frac{n_+-n_-}{n_++n_-}=\frac{n_+-n_-}{n_e},\label{xi}
\end{equation}
where $n_+$ and $n_-$ denote the number density of spin-up and spin-down electrons,
respectively~\footnote{Here spin-``up'' and spin-``down'' have only relatively meanings.
We define the electrons with spin-``up'' whose intrinsic magnetic moments
have the same direction as the external magnetic fields.}.
We find that, in solid quark-cluster stars, the value of $\xi$ could be non-zero (although
are some tiny values), which means
that the star has non-zero net macroscopic magnetic moment.
In some cases, the corresponding magnetic moment
per unit mass $\mu_0$ could be higher than $10^{-4}$ Gauss cm$^3$ g$^{-1}$,
which is large enough to induce the observed  magnetic fields of pulsars with $B\sim 10^{11}-10^{13}$ Gauss.
It should be clarified that, in this paper we will not explain the origin of super strong magnetic fields of the so called ``magnetars'', whose dipole magnetic fields are though to be larger than $10^{14}$ Gauss.
It is worth noting that, in the framework of quark-cluster star model, the gravitational and elastic energies could be large enough to account for the observed energy releasing~\cite{xty2006,Lai:2008cw,zhou2014}, so the super strong magnetic fields are (e.g., in the popular magnetar model) unnecessary.

In fact, in condensed matter physics, a full and complete description of ferromagnetism of electron system is still very complex and has not achieve a satisfying stage.
A microscopic calculation from first principle is certainly very difficult and is not the focus of this paper, but the leading order approximation in our simple model shows that the magnetism of solid quark-cluster matter could be possible.

This paper is arranged as follows. In \S II we will show the basic properties of
quark-clusters and electrons in solid quark-cluster stars.
Based on the solid quark-cluster star model,  in \S III we will demonstrate
the spontaneous magnetization of electrons in solid quark-cluster stars, using a toy model and then
a more quantitative model, and show that the induced magnetic fields could be significant.
Conclusions and discussions are made in \S IV.

\section{Quark-clusters and electrons in solid quark-cluster stars}

The average baryon number density $n_b$ of a pulsar-like compact star is about $3n_0$,
where $n_0$ is the baryon number density of saturate nuclear matter.
We have proposed that pulsar-like compact stars could be quark-cluster stars
~\cite{Lai:2008cw, LX2009b},
because the strongly interacting quarks could be grouped into quark-clusters
~\cite{Xu03, Xu:2010}.
At low enough temperatures, quark-clusters could crystallize into solid state, just like
the phase transition of normal matter from liquid to solid states.
The number of quarks in each quark-cluster, $N_q$, is seen as a free parameter which
is related to the interaction details.
Based on the symmetric consideration, the most possible value of $N_q$ could be 18, which means
that quark-clusters are singlets of spins, flavors and colors.
Certainly the choice of $N_q$ could have many possibilities, e. g. the $H$-cluster stars,
composed of $H$-dibaryons (an $H$-dibaryon is the bound state of two $\Lambda$-particles),
was proposed as a kind of quark-cluster stars.~\cite{lgx2013}, in which case $N_q=6$.
The constrains of $N_q$ by the maximum mass of pulsars were also be studied~\cite{Lai:2010wf},
and we find that the 2$M_\odot$ pulsar PSR 1614-2230~\cite{Demorest:2010bx}
infers $N_q\leqslant 10^3$.

A quark star could be considered as a gigantic nucleus, with electrons inside,
but changing from two-flavor ($u$, $d$) to three-flavor ($u$, $d$ and $s$) symmetries
~\cite{Xu03}.
The $H$-cluster star model was proposed based on three-flavor symmetries~\cite{lgx2013}.
If the star is composed of equal numbers of $u$, $d$ and $s$ quarks, no electrons
will exist.
In this case, the three-flavor symmetry may result in a ground state of matter,
as Bodmor-Witten conjecture said~\cite{Bodmer:1971we, Witten:1984rs}.
The mass difference between $u$, $d$ and $s$ quarks would break the symmetry,
but on the other hand, the interaction between quarks would lower the effect
of mass difference
and try to restore the symmetry.
The amount of symmetry breaking could be estimated based on perturbative calculations
~\cite{farhi-jaff1984}, which found the ratio of number densities of electrons to that of baryons
$n_e/n_b=Y_e$ might be smaller than $10^{-4}$.
Although it is difficult for us to calculate how strong the interaction between quarks
is, the non-perturbative nature and the energy scale of the system make it reasonable
to assume that the degree of the light flavor symmetry breaking is small.
In the following calculations, we assume $Y_e$ is ranged from $10^{-6}$ to $10^{-4}$.
Because of small $Y_e$, most of quark-clusters are electric neutral, but the quarks
inside them have electric charge.

Let us then show why the electrons in quark-cluster stars are completely itinerant.
The wave number of electrons with number density $n_e$ at Fermi surface is
\begin{equation}
k_F=(3\pi^2n_e)^{\frac{1}{3}} \simeq \frac{10\ {\rm MeV}}{\hbar c}\cdot
\left(\frac{Y_e}{10^{-5}}\right)^{\frac{1}{3}}\left(\frac{n_b}{3n_0}\right)^{\frac{1}{3}},
\label{kF}
\end{equation}
so electrons are relativistic, with Fermi energy of about 10 MeV.
The Coulomb attraction by quark-clusters on electrons is
\begin{equation}
E_c=\alpha \cdot n_e^{\frac{1}{3}}\simeq 10^{-2}\ {\rm MeV}\cdot
\left(\frac{Y_e}{10^{-5}}\right)^{\frac{1}{3}}\left(\frac{n_b}{3n_0}\right)^{\frac{1}{3}},
\label{Ec}
\end{equation}
($\alpha=1/137$ is the fine structure constant)
which means the Coulomb attraction cannot bound electrons because
of the kinetic energy is much larger than the binding energy, and all the the electrons
are itinerant.
This is quit different from the electrons in normal solid.

In summary, the differences from the electrons in solid quark-cluster stars and normal solid are
at least the following two aspects:
(1) the electrons are relativistic in solid quark-cluster stars and non-relativistic in normal solid;
and
(2) in solid quark-cluster stars all of the electrons are nearly freely moving around
rather than bond by the lattices.
%%We proposed that, pulsar-like compact stars could be quark-cluster stars,

\section{Spontaneous magnetization of electrons}

In solid quark-cluster stars, electrons repulse each other because of Coulomb interaction.
Electrons tend to have balanced spins to save kinetic energy, which is the case for
ideal (non-interacting) Fermi gas; and on the other hand, they also tend to have
unbalanced spins to save interaction energy, if Coulomb interaction is taken into
account.
Therefore, there is a competition in strongly degenerate electron gas, and in this paper
we are focusing on this issue.
If $\xi\neq 0$,
then there are net macroscopic magnetic moment.
The star could be composed of many magnetic domains, which has net
macroscopic magnetic moments due to the unbalanced spins.
Applying an external magnetic field (e.g. a fossil field), the directions of
magnetic domains will tend to align, and the maximum magnetic field will
achieve when all of the magnetic domains align perfectly.

In the following, we will first show a toy model which qualitatively give the
amount of unbalanced spins which could induce the magnetic fields of pulsars,
and then we will show a more quantitative way to demonstrate this.

\subsection{A toy model}

A pulsar with a dipole magnetic field $B\sim 10^{12}$ Gauss and radius $R\sim 10$ km
has the magnetic dipole moment $\sim 5\times 10^{29}\ \rm Gauss \cdot cm^3$.
If the baryon density $n_b=3n_0$ and the ratio of number densities of electrons to that of baryons $Y_e=10^{-5}$,
the total number of electrons is about $10^{52}$.
We then try to estimate if the the unbalanced spins of electrons could  account for
such magnetic moment.
Because the electrons with unbalanced spins will save interaction energy,
the electrons on the Fermi surface, within the momentum depth of $E_c/c$,
will tend to have the same spin.
From Eqs.(\ref{kF}) and (\ref{Ec}), the total number of such electrons is
\begin{equation}
\frac{4\pi p_F^2 E_c/c}{4\pi p_F^3/3}\sim 10^{49},
\end{equation}
which is just the number of unbalanced electrons, and
we can see that the amount of unbalanced spins $\xi\sim 10^{49}/10^{52}\sim 10^{-3}$.
The star will have a maximum magnetic dipole moment $\mu_{\rm d}$
if all of the magnetic moments of $10^{49}$ electrons align perfectly
\begin{equation}
\mu_{\rm d}\simeq \frac{4\pi}{3}\mu_e n_e \xi R^3
\sim 2\times 10^{29}\ \rm Gauss\ cm^3,
\end{equation}
where $\mu_e=9\times 10^{-21}\ \rm Gauss \cdot cm^3$ is the Bohr
magneton of electrons.
As a highly degenerate system, when the Coulomb interaction is taken into account,
the amount of electrons with the same only constitutes a tiny fraction of the whole electrons,
and the qualitative estimation convinces us that it could be enough to account for the
origin of strong magnetic fields of pulsars.

\subsection{Spin-alignment of electrons}
\label{subsection_spin}

Coulomb repulsion is responsible to the spontaneous magnetization of elections,
as indicated above.
Now we will demonstrate this in a more quantitative way.
In fact, it is generally believed in condensed matter physics that, a dilute Fermi gas
with repulsive interactions can undergo a ferromagnetic phase transition to a
 state with unbalanced spins, which will happen when the number density of electrons
reaches a critical value.
To show this, one could then treat the Coulomb interaction between two electrons to be simplified
as~\cite{huang}
\begin{equation}
v_{ij}=C\delta(\vec{r_i}-\vec{r_j}),\label{Vij}
\end{equation}
where $C$ is related to the scattering length $a$ and the mass of electrons $m$ via
~\cite{kaplan}
\begin{equation}
C=\frac{4\pi a}{m}.
\end{equation}
Although the above relation between $C$ an $a$ is come from low energy scattering
of electrons, we assume that it could be extrapolate to high energy case.

It should be noted that the condition for simplifying Coulomb interaction to
Delta function is that $k_F\cdot a\ll 1$, where $k_F$ is the wave number of particles at
Fermi surface, whereas in condensed matter physics,
where the electrons are treated as degenerate
non-relativistic Fermions, the condition for ferromagnetism is
$k_F\cdot a>\pi/2$~\cite{huang}.
That is to say, the model constructed based on Eq.(\ref{Vij}) can only serve as an instructive
way to show how the repulsion between Fermions to enhance the spin alignment.

However, for the electrons in a solid quark-cluster star, the condition $k_F\cdot a \ll 1$ would
be satisfied.
The long-range Coulomb interaction would change into short-ranged, due to screening effect in polarized quark-cluster matter.
Quark-clusters could be polarized in the presence of electrons, with polarization much larger than that of vacuum.
The distance between two neighboring quark-cluster $d\sim 2\ {\rm fm}\ (n_b/3n_0)^{-1/3}(N_q/18)^{1/3}$, and this could be comparable with the size of each quark-cluster $l_q$.
The individual quark inside each quark-clusters has electric charges, so the electromagnetic interaction between quarks that belong respectively to two neighboring quark-clusters could change the distribution of quarks inside both clusters.
Once an electron near a quark-cluster change the distribution of quarks inside it, the arrangement of electric charge inside quark-cluster would spread out by a small but significant amount because $d\sim l_q$, although the strength of electromagnetic interaction is weaker than that of strong interaction by 2 orders of magnitude.
This would lead to polarizability $\varepsilon$ which is much larger than the polarizability of vacuum $\varepsilon_0$.
That means, to get significant Coulomb interaction, an electron should be very close to another electron, with distances much smaller than the average distance $\sim n_e^{-1/3}$.
Unfortunately it is now difficult to calculate qualitatively the polarizability of quark-cluster matter, but we can make the approximation that the values of scattering length of electrons $a=\eta \cdot d$ where $1\leqslant\eta \leqslant10$, then
\begin{equation}
k_F\cdot a\simeq 0.4\left(\frac{\eta}{10}\right)\left(\frac{Y_e}{10^{-5}}\right)^{1/3} \left(\frac{N_q}{18}\right)^{1/3}.\label{kfa}
\end{equation}
We will see in below that, for parameter space which give large enough magnetic momentum, the condition making the Stoner model valid, $k_F\cdot a\ll 1$, will be satisfied.

The corresponding interaction energy for electron system composed of $N$ electrons is
\begin{equation}
E_{\rm int}=\left(\phi, \sum_{i<j}v_{ij}\phi \right)=\frac{4\pi a}{m}\frac{N_+ N_-}{V},
\end{equation}
where $\phi$ is the wave function of electron system,
$N_+$ and $N_-$ are the total number of spin-up and spin-down electrons respectively,
$N=N_++N_-$, and $V$ is the volume of the system.
The total energy density of relativistic electrons,
taking into account the kinetic energy and interaction energy, is then
%\begin{eqnarray}
%\epsilon&=&\frac{3}{8}(3\pi^2)^{\frac{1}{3}}n^{\frac{4}{3}}\left[(1+\xi)^{\frac{4}{3}}+(1-\xi)^{\frac{4}{3}}\right]
%\nonumber\\&+&\frac{\pi a}{m} n^2(1-\xi^2)
%\end{eqnarray}
\begin{equation}
\epsilon=\frac{3}{8}(3\pi^2)^{\frac{1}{3}}n_e^{\frac{4}{3}}
\left[(1+\xi)^{\frac{4}{3}}+(1-\xi)^{\frac{4}{3}}\right]
+\frac{\pi a}{m} n_e^2(1-\xi^2).\label{epsilon}
\end{equation}
The condition for ferromagnetism is that a value of $\xi$ in the range $(0,1]$
minimizes $\epsilon$.
Inserting the definition of $N_q$ and $Y_e$, and assuming that $a\simeq d$, this condition can be written as
\begin{equation}
N_q\cdot Y_e^2>\frac{m^3}{24\pi\cdot n_b}=4.8\times 10^{-10}
\left(\frac{n_b}{3n_0}\right)^{-1}, ~\label{condition}
\end{equation}\label{Ye}
which means that larger $N_q$ or larger $Y_e$ would favor ferromagnetism.
With non-zero $\xi$ that minimize the energy density in Eq.(\ref{epsilon}), the electron
system will have non-zero macroscopic magnetic moment under an external field.

To derive the total magnetic moment of the electron system, we can define the
magnetic moment per unit mass $\tilde{\mu}_0$,
so the total magnetic moment $\mu$ of the a star with mass $M$ is $\mu=\tilde{\mu}_0M$,
or
\begin{eqnarray}
\mu&=&3\times 10^{29}\ {\rm Gauss\ cm^3\ g^{-1}}\nonumber\\
&\times&\left(\frac{\tilde{\mu}_0}{10^{-4}\ \rm Gauss\ cm^3\ g^{-1}}\right)
\left(\frac{M}{1.5\ M_\odot}\right).  ~\label{mu}
\end{eqnarray}
Then we define the magnetic moment per unit mass $\mu_0$,
when all the magnetic moments
of electrons point to the same direction, so $\mu_0$ is proportional to $\xi$
that minimize the energy density $\epsilon$ in Eq.(\ref{epsilon}).
%where $\tilde{\mu}_0$ is the real magnetic moment per unit mass of a star.
%
Like normal ferromagnetic material, the ferromagnetic quark-cluster stars should be composed
of magnetic domains.
The direction of each domain would not be completely the same with each other, so
the real magnetic moment per unit mass of a star $\tilde{\mu}_0$ would be smaller
than $\mu_0$.
The relation between $\tilde{\mu}_0$ and $\mu_0$ depends on many uncertain
factors such as the configurations of magnetic fields and the shapes of domains, so it is
difficult for us to derive an accurate expression.
Here we assume that under an external field, the degree of alignment is sufficient, and
$\tilde{\mu}_0\sim 0.01-0.1\mu_0$.
So if $\mu_0\geqslant 10^{-3}\sim 10^{-2}$ Gauss cm$^3$ g$^{-1}$, then $\tilde{\mu}_0\geqslant 10^{-4}$ Gauss cm$^3$ g$^{-1}$, corresponding to $B\geqslant 10^{12}$ Gauss.

Figure~\ref{fig_mu} shows the relation between $\mu_0$ and $Y_e$, in two cases $\eta=1$ (thick blue lines) and $\eta=10$ (thin black lines), and we assume a homogeneous star with number density $n_b=3n_0$.
Solid, dashed and dash-dotted lines correspond to $N_q=$6, 18, 100 respectively.
From this figure we can see that,
if $Y_e\geqslant10^{-7}$, the corresponding $\mu$ could be large enough to account for
the magnetic moments of pulsars.
For example, $Y_e\simeq 10^{-6}$ can give satisfying value of magnetic moment
of a pulsar with magnetic field $B=10^{12}$ Gauss and radius $R=10$ km, if $\eta\simeq10$ and
$\tilde{\mu}_0\sim 0.01\mu_0$, as shown in Eq.(\ref{mu}).
In the case $\eta=1$, if $Y_e<2\times10^{-6}$, spontaneous magnetization is
unlikely to happen, consistent with condition~(\ref{condition}).
Moreover, in this figure we can see that, for parameter space which give large enough magnetic momentum the condition making the Stoner model valid, $k_F\cdot a\ll 1$, will be satisfied, as shown in Eq(\ref{kfa}).
The quick drops of curves in the bottom-left result from the fact that, given a value of $N_q$,
both of $\xi$ that minimizes energy density~(\ref{epsilon}) and the number density of electrons
$n_e$ decrease with $Y_e$.

\begin{center}
\includegraphics[width=9cm]{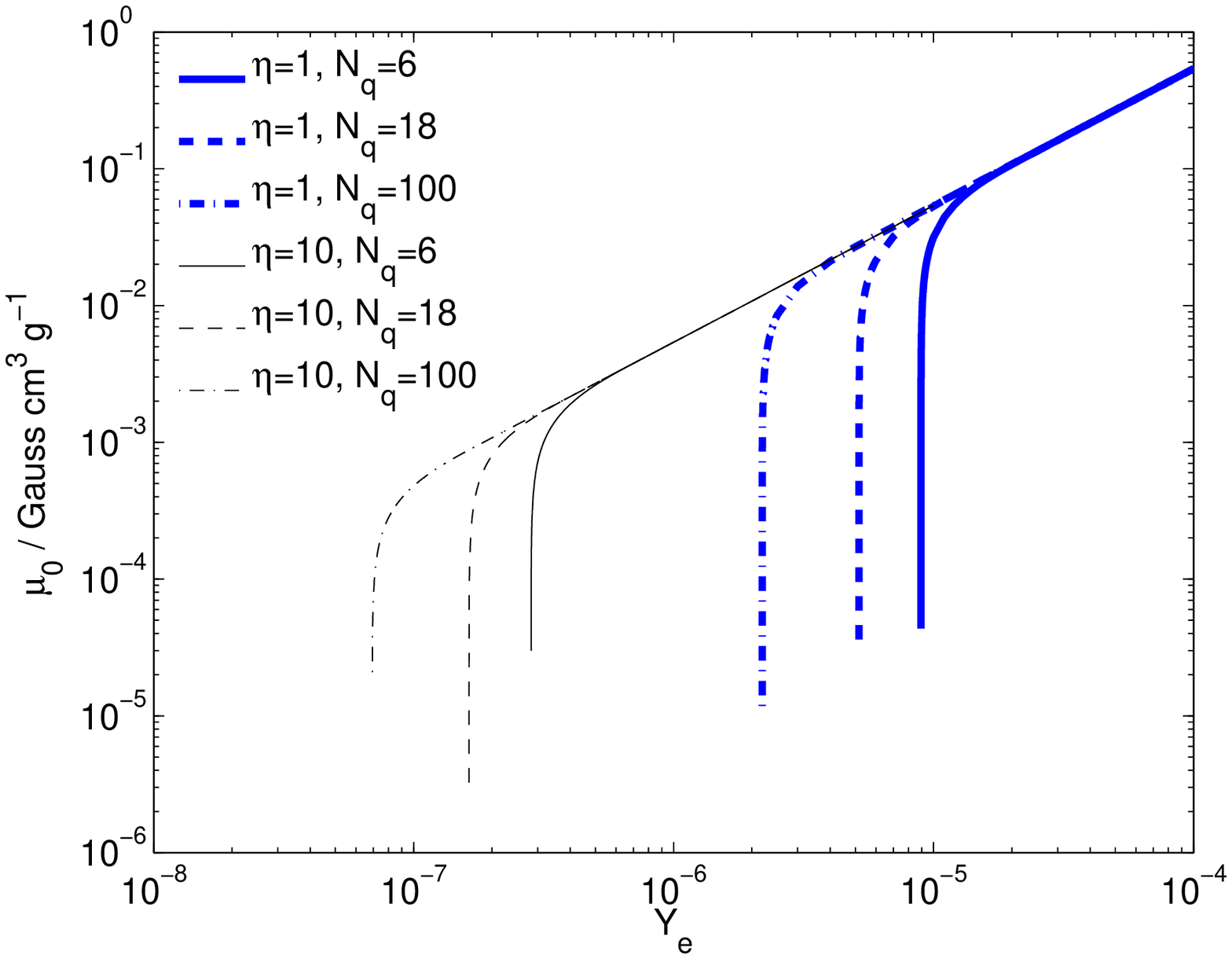}
\figcaption{\label{fig_mu} The relation between $\mu_0$ (magnetic moment per unit mass) and $Y_e$ (ratio of number densities of electrons to that of baryons $n_e/n_b$), in two cases $\eta=1$ (thick blue lines) and $\eta=10$ (thin black lines), where $\eta$ denotes the strength of polarization, defined as the ratio of electrons' scattering length $a$ and the distance between two nearby quark-clusters $d$.
Solid, dashed and dash-dotted lines correspond to $N_q=$6, 18, 100 respectively.
If $\mu_0\geqslant 10^{-3}\sim 10^{-2}$ Gauss cm$^3$ g$^{-1}$, then $\tilde{\mu}_0\geqslant 10^{-4}$ Gauss cm$^3$ g$^{-1}$, corresponding to $B\geqslant 10^{12}$ Gauss. }
\end{center}

A star should have different densities from center to surface.
Higher baryon densities means weaker interaction and consequently larger $Y_e$,
so from this figure and the relation~(\ref{condition}) we can see that,
at higher baryon densities the condition for spontaneous magnetization
would be less strict.
Because the range of $n_b$ in pulsars is about from $2n_0$ to $10n_0$,
the spontaneous magnetization could occur in almost all of the region in the star, leading to
a sufficiently large macroscopic magnetic moment.
We can see that, in the model of spontaneous magnetization of electrons we show in this paper, there is large enough parameter-space for $Y_e$ which can lead to magnetic fields with strength $B\sim 10^{11}$ Gauss to $\sim 10^{13}$ Gauss.

\section{Conclusions and discussions}

In this paper we demonstrate that the strong magnetic fields of pulsars
could be originated from the spontaneous magnetization of electrons
in solid quark-cluster star model.
Due to the relatively low densities of electrons compared with that of quark-clusters, and polarization of quark-cluster matter in compact stars, the screening effect between electrons could be significant, that is to say,
the scattering length of electrons could be much smaller than the average distance between electrons.
Under such situation, the Coulomb interaction between electrons could be simplified as
a $\delta$ function in Eq.(\ref{Vij}), and subsequently we can apply the Stoner's model
to demonstrate ferromagnetism.
The competition between Coulomb repulsion and the Pauli exclusion principle could make the
electron-system with unbalanced spins more stable than that with balanced spins.

The amount of unbalanced spins $\xi$ depends on $Y_e$.
Larger $Y_e$ results in larger $\xi$ and consequently induce larger
magnetic moment per unit mass $\tilde{\mu}_0$.
Note that $\tilde{\mu}_0<\mu_0$, where $\mu_0$ is the value of $\tilde{\mu}_0$ when all of the magnetic
domains point to the same direction.
If $Y_e>10^{-7}$, the corresponding $\mu$ could be large enough to account for
the magnetic moments of pulsars with $B\geqslant 10^{12}$ Gauss.
If $Y_e<10^{-7}$, spontaneous magnetization is
unlikely to happen.
Therefore we show that the spontaneous
magnetization of electrons could account for the origin of strong magnetic
fields of pulsars.
If the strengths of magnetic fields of compact stars are generally in the range $B\simeq 10^{12}$ Gauss,
we can infer that the allowed values of $Y_e$ are in the range from $\sim 10^{-7}$ to $\sim 10^{-5}$.
This could hint quantitatively the symmetry broken of three light flavors in strange quark-cluster matter.

From the model discussed in this paper, we could also infer that the strength of magnetic fields of pulsars
with same mass and radius should be the same, because of the same total magnetic momentum.
As we have shown above, if the value of $Y_e$ is in the range from $\sim 10^{-7}$ to $10^{-5}$, then the strength of magnetic field would be in the range from $\sim 10^{11}$ to $10^{13}$ Gauss.
However, the magnetic fields of some pulsars, especially the millisecond pulsars, are as low as
$\sim 10^8$ Gauss.
So how to explain such weak magnetic fields?
Almost all of the discovered millisecond pulsars are in binaries, and one possible mechanism for
reducing the strength of dipole fields is related to the accretion process~\cite{zhang1998}.
The accreted material could squeeze some of the surface material towards the equator and eventually
bury the magnetic fields at the equator.
The strength of dipole magnetic fields could be decreased rapidly and would reach a minimum value
$\sim 10^8$ Gauss.

Another kind of pulsar-like compact stars includes AXPs (anomalous X-ray pulsars) and SGRs
(soft gamma-ray repeaters).
One proposal to solve the energy budget is that they are highly-magnetized pulsars, i.e. the so-called
magnetars, with the strength of magnetic fields $\geqslant 10^{14}$ Gauss.
However, a rigid quark-cluster star can also provide free energies being high enough for bursts and even giant flares, in terms of
gravitational and elastic energies~\cite{xty2006,Lai:2008cw}, and the super strong magnetic fields are thus
unnecessary.
The ``low'' dipole field ($\leqslant 7.5\times 10^{12}$ Gauss) of SGR 0418+5729~\cite{rea2010}
has challenged magnetar
model, and more challenge of magnetar model can be found in~\cite{tx2011} and references therein.
Therefore, in this paper we assume that all of the pulsar-like compact stars (at lease at the moment of solidification)
have magnetic fields with the strength $B\leqslant 10^{13}$ Gauss.

In the model of quark-cluster stars, the values of $N_q$ and $Y_e$ depend
on the properties of strong interaction at low energy scale.
In our previous work, we constrained $N_q$ by the maximum mass of pulsars,
and in this paper we give constrains on $Y_e$ based on the proposal that
the strong magnetic fields of pulsars are originated from the spontaneous
magnetization of electrons.
Smaller $Y_e$ implies stronger interaction between quarks, since if interaction is strong,
the effect of mass difference between $u$, $d$ and $s$ would become
less significant and make the number of $s$ quarks to be larger.
So this picture could tell us that,
although the interaction between quarks are strong enough to group quarks into
clusters, it may not enough to make $Y_e$ to be smaller than about
$10^{-7}$ in order to produce enough strong magnetic fields for pulsars.

Besides magnetic properties, electrons are also important for the radiation properties of pulsars.
In quark star models, the stars may be enveloped in thin electron layers which uniformly
surround the entire star.
The hydrocyclotron oscillation of electron layers could explain the observed absorption
features of some pulsars~\cite{xbw2011}.

Some properties of strange matter should be modified by strong magnetic fields.
In our previous papers about quark-cluster stars, however, we neglected the impacts of
magnetic fields on the equation of state,
and took the whole star as non-magnetic.
The reason is that, comparing the orders of magnitude, we can see that
the energy density of magnetic fields $B^2/8\pi \sim 10^{23}
{\rm erg/cm^3}\cdot(B/10^{12}
{\rm Gauss})^2$ is much smaller than the energy density of rest mass of strange matter
$\sim 10^{35} {\rm erg/cm^3}\cdot (n_b/3n_0)$.
The global features of a strange star, such as mass and radius, would consequently not
change significantly if magnetization is included.
In this paper, for the same reason, we still do not discuss the impacts of
magnetic fields on the state of strange matter.
The effects of strong magnetic fields on properties of strange matter, such as
the bulk energy density and transport properties, are interesting topics, and
deserve detailed studies.
Anyway, though we think this is out of the scope of this paper, we will study more about
this in the future.

It should also be noted that, the properties of relativistic electrons are in fact not
very certain to us.
Limited by the model used in normal solid containing non-relativistic electrons,
we are in lack of the knowledge about the relativistic effects on electrons
moving in solid quark-cluster stars, including the scattering length and the influences of
lattices on electrons.
On this point of view, the model we present in this paper for relativistic electron-system
is to some extent an approximation to the electrons in quark-cluster stars.
The spin-unbalance of electrons could give rise the the strong magnetic fields of pulsars,
but the detailed model to demonstrate this based on more reliable ground remains
to be constructed.
\\

\acknowledgments{We thank Profs. Wei Guo, Guangshan Tian and Lan Yin in PKU for useful discussions.
This work is supported by the 973 program (No. 2012CB821801),
the National Natural Science Foundation of China (11203018, 11225314, 11365022),
the West Light Foundation (XBBS-2014-23),
the Science Project of Universities in Xinjiang (XJEDU2012S02) and
the Doctoral Science Foundation of Xinjiang University (BS120107).}

\end{multicols}

\vspace{10mm}

\vspace{-1mm}
\centerline{\rule{80mm}{0.1pt}}
\vspace{2mm}

\begin{multicols}{2}

\end{multicols}

\clearpage

\end{CJK*}
\end{document}